\documentclass[12pt]{article}

\newcommand{\pslash}{\not \! p}
\newcommand{\kslash}{\not \! k}

\usepackage{amsmath,amssymb,amscd,amsbsy,amsgen,amsopn,amstext,
amsxtra}
\usepackage{hyperref}
\usepackage{cite}
\usepackage{graphicx}

\begin{document}

\vskip 0.5 truecm

\begin{center}
{\Large{\bf Neutrino-antineutrino mass splitting in the Standard Model and 
baryogenesis
}}
\end{center}
\vskip .5 truecm
\begin{center}
{\bf { Kazuo Fujikawa$^\dagger$ and Anca Tureanu$^*$}}
\end{center}

\begin{center}
\vspace*{0.4cm} {\it {$^\dagger$ Mathematical Physics Laboratory, RIKEN Nishina Center,\\
Wako 351-0198, Japan\\
 $^*$Department of Physics, University of
Helsinki, P.O.Box 64, \\FIN-00014 Helsinki,
Finland\\
}}
\end{center}


\begin{abstract}
On the basis of a previously proposed mechanism of neutrino-antineutrino mass splitting in the Standard Model, which is Lorentz and $SU(2)\times U(1)$ invariant but non-local to evade $CPT$ theorem, we discuss the possible implications of neutrino-antineutrino mass splitting on neutrino physics and baryogenesis. It is shown that non-locality within a distance scale of the Planck length, that may not be fatal to unitarity in generic effective theory, can generate the neutrino-antineutrino mass splitting of the order of observed neutrino mass differences, which is tested in oscillation experiments, and non-negligible baryon asymmetry depending on the estimate of sphaleron dynamics. The one-loop order induced electron-positron mass splitting in the Standard Model is shown to be finite and estimated  at $\sim 10^{-20}$ eV, well below the experimental bound $< 10^{-2}$ eV. The induced $CPT$ violation in the $K$-meson in the Standard Model is expected to be even smaller and well below the experimental bound $|m_{K}-m_{\bar{K}}|<0.44\times 10^{-18}$ GeV. 

\end{abstract}

 
\section{Introduction}
The baryogenesis on the basis of $CPT$ violating proton-antiproton mass splitting~\cite{zeldovich} and neutrino-antineutrino mass splitting~\cite{barenboim} has been discussed some time ago. The possible neutrino-antineutrino mass splitting has also been discussed in connection with the phenomenology of neutrino oscillation~\cite{pakvasa, murayama}. On the other hand, an experimental study~\cite{adamson2} states that  "we have detected no evidence for Lorentz invariance violation in the antineutrino data set." 
It has been recently shown that the neutrino-antineutrino mass splitting can be realized in the Standard Model by preserving both Lorentz invariance and $SU(2)\times U(1)$ gauge symmetry~\cite{CFT1}. This model is based on a Lorentz invariant $CPT$ violation mechanism which also breaks $C$ and $CP$ but preserves $T$~\cite{CFT2}. The basic ingredient to evade the $CPT$ theorem~\cite{pauli} in this scheme is that the theory is assumed to be non-local~\cite{CDNT, JGB}, although the original scheme of $CPT$ violation with $T$ violation but preserving $C$ and $CP$ does not generate particle-antiparticle mass splitting~\cite{CDNT}. 
In this paper, we perform a detailed evaluation  of this neutrino-antineutrino mass splitting in the Standard Model and discuss  its possible implications on neutrino physics and baryogenesis. It is shown that the non-locality within a distance scale of the Planck length, which may not be fatal to unitarity in  generic effective theory, can generate the neutrino-antineutrino mass splitting of the order of observed neutrino mass differences and non-negligible baryon asymmetry, depending on the estimate of sphaleron dynamics. The  one-loop order electron-positron mass splitting induced by the neutrino-antineutrino mass splitting in the Standard Model is confirmed to be finite and kept well below the experimental bound.

\section{The model} 
We start with a minimal extension of the Standard
Model~\cite{weinberg} by incorporating the right-handed neutrino
\begin{equation}
  \psi_{L}=\left(
  \begin{array}{c}
   \nu_{L}\\ e_{L}
  \end{array}
  \right), \ \quad
   e_{R},\ \nu_{R} .\label{(1)}
\end{equation}
For simplicity, we consider only a single flavor of leptons. The part of the Standard Model Lagrangian relevant to our discussion
is given by
\begin{eqnarray}\label{standard}
{\cal L}&=&i\overline{\psi}_{L}\gamma^{\mu}
\left(\partial_{\mu} - igT^{a}W_{\mu}^{a}
             - i\frac{1}{2}g^{\prime}Y_{L}B_{\mu}\right)\psi_{L}
\nonumber\\
         & +&i\overline{e}_{R}\gamma^{\mu}(\partial_{\mu}
             + ig^{\prime}B_{\mu})e_{R}
+i\overline{\nu}_{R}\gamma^{\mu}\partial_{\mu}\nu_{R}\nonumber
\\
            &+&\left[ -
\frac{\sqrt{2}m_{e}}{v}\overline{e}_{R}\phi^{\dagger}\psi_{L}
-\frac{\sqrt{2}m_{D}}{v}\overline{\nu}_{R}\phi_{c}^{\dagger}\psi_{L}
             -\frac{m_{R}}{2}\nu_{R}^{T}C\nu_{R} + h.c.\right],
\end{eqnarray}
with $Y_{L}=-1$, and the Higgs doublet and its $SU(2)$ conjugate
\begin{eqnarray}\label{(3)}
  \phi=\left(
  \begin{array}{c}
   \phi^{+}\\ \phi^{0}
  \end{array}
  \right), \ \ \ \ \ \
   \phi_{c}\equiv i\tau_{2}\phi^{\star}=\left(
  \begin{array}{c}
   \bar{\phi}^{0}\\ -\phi^{-}
  \end{array}\right).
\end{eqnarray}
The operator $C$ stands for the charge conjugation matrix for
spinors. The term with $m_{R}$ in the above Lagrangian is the
Majorana mass term for the right-handed neutrino.
We tentatively assume that the right-handed Majorana mass
vanishes, namely, we adopt the Dirac limit $m_{R}=0$ with enhanced lepton number conservation. Physically, this assumption amounts to the ansatz that {\em all the masses arise from the Higgs boson}, which has been discovered recently. Related schemes of neutrino masses have been discussed by various authors in the past~\cite{wolfenstein, langacker, extra-dim} (for reviews, see \cite{rev}).

Our next observation is that the combination
\begin{eqnarray}\label{(4)}
\phi_{c}^{\dagger}(x)\psi_{L}(x)
\end{eqnarray}
is invariant under the full $SU(2)_{L}\times U(1)$ gauge symmetry. One
may thus add a hermitian non-local Higgs coupling with a real parameter $\mu$ to the Lagrangian  \eqref{standard},
\begin{eqnarray}\label{(5)}
{\cal L}_{CPT}(x)&=&-i\frac{2\sqrt{2}\mu}{v}\int
d^{4}y\Delta_{l}(x-y)\theta(x^{0}-y^{0})\{\bar{\nu}_{R}(x)\left(\phi_{c}^{\dagger}(y)\psi_{L}(y)\right)\nonumber\\
&&\hspace{3 cm}-\left(\bar{\psi}_{L}(y)\phi_{c}(y)\right)\nu_{R}(x)\},
\end{eqnarray}
without spoiling the basic Lorentz invariance and the $SU(2)_{L}\times U(1)$ gauge symmetry. Here we defined
\begin{eqnarray}\label{(6)}
\Delta_{l}(x-y)\equiv \delta\left((x-y)^{2}-l^{2}\right)-\delta\left((x-y)^{2}-(l^{\prime})^{2}\right)
\end{eqnarray}
with $l$ and $l^{\prime}\neq l$ standing for fixed length scales. This factor differs from $\delta((x-y)^{2}-l^{2})$ used in~\cite{CFT1}, and this modified $\Delta_{l}(x-y)$ avoids a quadratic infrared divergence, as is explained later. In our concrete evaluation below, we set $l^{\prime}=0$ for simplicity, although a very small $l^{\prime}\ll l$ may help to make the separation of the future and past light-cones clear without modifying the essence of our analysis.

In the unitary gauge, $\phi^{\pm}(x)=0$ and $\phi^{0}(x)\rightarrow (v
+ \varphi(x))/\sqrt{2}$, the neutrino mass  term (with $m_{R}=0$)
becomes in terms of the action
\begin{eqnarray}\label{mass}
S_{\nu \rm mass}&=&\int
d^{4}x\Big\{-m_{D}\bar{\nu}(x)\nu(x)\left(1+\frac{\varphi(x)}{v}\right)\\
&& -i\mu\int
d^{4}y\Delta_{l}(x-y)\theta(x^{0}-y^{0})\nonumber\\
&&\times \left[\bar{\nu}(x)\left(1+\frac{\varphi(y)}{v}\right)(1-\gamma_{5})\nu(y)
-\bar{\nu}(y)\left(1+\frac{\varphi(y)}{v}\right)(1+\gamma_{5})\nu(x)\right]\Big\}
\nonumber\\
&=&\int d^{4}x\Big\{-m_{D}\bar{\nu}(x)\nu(x)\left(1+\frac{\varphi(x)}{v}\right)\nonumber\\
&& -i\mu\int
d^{4}y\Delta_{l}(x-y)\left[\theta(x^{0}-y^{0})-\theta(y^{0}-x^{0})\right]
\bar{\nu}(x)\nu(y)\nonumber\\
&&+i\mu\int
d^{4}y\Delta_{l}(x-y)\bar{\nu}(x)\gamma_{5}\nu(y)\nonumber
\\
&&-i\frac{\mu}{v}\int
d^{4}y\Delta_{l}(x-y)\theta(x^{0}-y^{0})
\left[\bar{\nu}(x)(1-\gamma_{5})\nu(y)-\bar{\nu}(y)(1+\gamma_{5})\nu(x)\right]\varphi(y)\Big\}, \nonumber
\end{eqnarray}
where we have changed the naming of integration variables 
$x\leftrightarrow y$ in some of the terms and used
$\theta(x^{0}-y^{0})+\theta(y^{0}-x^{0})=1$. The term 
\begin{eqnarray}\label{(8)}
-i\mu\int
d^{4}x\int d^{4}y\Delta_{l}(x-y)\left[\theta(x^{0}-y^{0})-\theta(y^{0}-x^{0})\right]
\bar{\nu}(x)\nu(y)
\end{eqnarray}
in the action  preserves $T$ but has $C=CP=CPT=-1$ and thus gives rise to particle-antiparticle mass splitting~\cite{CFT2}.

The equation of motion for the free neutrino is given by 
\begin{eqnarray}\label{(9)}
i\gamma^{\mu}\partial_{\mu}\nu(x)&=&m_{D}\nu(x)+i\mu\int
d^{4}y\Delta_{l}(x-y)\left[\theta(x^{0}-y^{0})-\theta(y^{0}-x^{0})\right]\nu(y)\nonumber\\
&-&i\mu\int d^{4}y\Delta_{l}(x-y)\gamma_{5}\nu(y).
\end{eqnarray}
By inserting an Ansatz for the solution,
$\psi(x)=e^{-ipx}U(p)$, into the equation of motion,
we obtain
\begin{eqnarray}\label{(2.6)}
\pslash U(p)&=&\Big\{m
+i[f_{+}(p)-f_{-}(p)]-i g(p^{2})\gamma_{5}\Big\}U(p),
\end{eqnarray}
where $f_{\pm}(p)$ is a Lorentz invariant quantity defined by
\begin{eqnarray}\label{(1.3)}
&&f_{\pm}(p)=\mu\int d^{4}z
e^{\pm ipz}\theta(z^{0})\left[\delta\left((z)^{2}-l^{2}\right)-\delta(z^{2})\right].
\end{eqnarray}
$f_{+}(p)$ and $f_{-}(p)$ are  inequivalent for  time-like $p$ due to the factor
$\theta(z^{0})$. The parity violating mass term in \eqref{(2.6)}, which is $C$ and $CPT$ preserving and thus does not contribute to the mass splitting, contains the factor
\begin{eqnarray}\label{(12)}
g(p^{2})=
\mu\int d^{4}ze^{ipz}[\delta((z)^{2}-l^{2})-\delta((z)^{2})].
\end{eqnarray}
The factor $f_{\pm}(p)$ is mathematically related to
the two-point Wightman function for a free scalar field. To be explicit,
\begin{eqnarray}\label{(13)}
\langle 0|\phi(x)\phi(y)|0\rangle=\int d^{4}p
e^{i(x-y)p}\theta(p^{0})\delta(p^{2}-m^{2}),
\end{eqnarray}
and the Wightman function is finite for $x-y\neq 0$ except for the possible cut
in the time-like separation, but divergent for the short distance $x-y\rightarrow 0$. This short distance behavior, whose leading term is mass independent, is related to the infrared $p\rightarrow 0$ behavior of $f_{\pm}(p)$ and $g(p^{2})$. Our modification $\Delta_{l}(x-y)=\delta((z)^{2}-l^{2})-\delta((z)^{2})$ thus eliminates the quadratic infrared divergence, which is independent of $l$, in $f_{\pm}(p)$ and 
$g(p^{2})$. Also, our mass splitting term is analogous to the discontinuity along the cut in the time-like separation $x-y$ of Wightman function.

For  time-like $p^{2}>0$, one may go to the frame where $\vec{p}=0$. Then the 
eigenvalue equation is given by 
\begin{eqnarray}\label{(14)}
p_{0}&=&\gamma^{0}[m_{D}-f(p_{0})-ig(p^{2}_{0})\gamma_{5}],
\end{eqnarray}
with
\begin{eqnarray}\label{(15)}
f(p_{0})&\equiv&-i[f_{+}(p_{0})-f_{-}(p_{0})]\nonumber\\
&=&4\mu\pi\int_{0}^{\infty}dz\Big\{\frac{z^{2}\sin [p_{0}\sqrt{z^{2}+l^{2}}]}{\sqrt{z^{2}+l^{2}}}-\frac{z^{2}\sin [p_{0}\sqrt{z^{2}}]}{\sqrt{z^{2}}}\Big\}
\end{eqnarray}
and 
\begin{eqnarray}\label{(16)}
g(p^{2}_{0})=4\mu\pi\int_{0}^{\infty}dz\Big\{\frac{z^{2}\cos [p_{0}\sqrt{z^{2}+l^{2}}]}{\sqrt{z^{2}+l^{2}}}-\frac{z^{2}\cos [p_{0}\sqrt{z^{2}}]}{\sqrt{z^{2}}}\Big\}.
\end{eqnarray}
For space-like $p^{2}<0$, one can confirm that the $CPT$ violating term vanishes, $f(p)=0$, by choosing $p_{\mu}=(0, \vec{p})$.

Since we are assuming that the $CPT$ breaking terms are small, we may
solve the mass eigenvalue equations iteratively by assuming that the
terms with
the parameter $\mu$, whose mass dimension is $[\mu]=[M]^{3}$, are much smaller than $m=m_{D}$. We then obtain
the mass eigenvalues of the neutrino and antineutrino at~\cite{CFT1}
\begin{eqnarray}\label{(17)}
m_{\pm}
&\simeq&m_{D}-i\gamma_{5} g(m_{D}^{2}) \pm
f(m_{D}).
\end{eqnarray}
The parity violating mass $-i\gamma_{5} g(m_{D}^{2})$ is now transformed away by a suitable
global chiral transformation without modifying the last term in \eqref{(17)}
to the order linear in the parameter $\mu$. In this way, the
neutrino-antineutrino mass splitting is incorporated in the
Standard Model by the Lorentz invariant non-local $CPT$ breaking
mechanism, without spoiling the $SU(2)_{L}\times U(1)$ gauge symmetry. The
Higgs particle $\varphi$ itself has a tiny $C$, $CP$ and $CPT$
violating coupling in \eqref{mass}.

\section{Evaluation of mass splitting}

We now explicitly evaluate the mass splitting in our model. As already mentioned, for space-like momentum $p^{2}<0$ in \eqref{(2.6)}, the
$CPT$ violating term vanishes, $f(p)=0$. We thus consider only the time-like momentum $p^{2}>0$ in the following
and use a generic notation $f(p)$.

With a uniform large cut-off $L$ of the spatial variable $z$ and after the change of the integration variable as
$y=\sqrt{z^{2}+l^{2}}$ in the first term of \eqref{(15)}, we have 
\begin{eqnarray}\label{(18)}
f(p)&=&4\pi\mu\Big\{\int_{l}^{\sqrt{L^{2}+l^{2}}}dy \sqrt{y^{2}-l^{2}}\sin [
p_{0}y]-\int_{0}^{L}dy y\sin [p_{0}y]\Big\}\nonumber\\
&=&-4\pi\mu[\theta(p_{0})-\theta(-p_{0})]\theta(p^{2})\Big\{\int_{l|p_{0}|}^{\sqrt{L^{2}+l^{2}}|p_{0}|}dv \frac{1}{[\sqrt{v^{2}-(l|p_{0}|)^{2}}+v]}\sin v\nonumber\\
&&+\int_{0}^{l}dy y\sin [
|p_{0}|y]-\int_{L}^{\sqrt{L^{2}+l^{2}}}dy y\sin [
|p_{0}|y]\Big\},
\end{eqnarray}
which clearly shows that the limiting value $f(\pm 0)=0$ for any fixed $L$. We here took into account the fact that $p^{2}\geq 0$. It is confirmed that the last term in \eqref{(18)} gives for large $L$
\begin{eqnarray}\label{(19)}
2\pi\mu l^{2}[\theta(p_{0})-\theta(-p_{0})]\theta(p^{2})\sin(|p_{0}|L),
\end{eqnarray}
which implies that the convergence of the integral  \eqref{(15)} is rather subtle. By ignoring the last term for the moment, the 
expression\eqref{(18)} is re-written for $|p_{0}|L\rightarrow\infty$
\begin{eqnarray}\label{(20)}
f(p)
&=&-4\pi\mu l^{2}[\theta(p_{0})-\theta(-p_{0})]\theta(p^{2})\\
&&\times\Big\{\int_{1}^{\infty}du \frac{1}{2u(\sqrt{u^{2}-1}+u)^{2}}\sin (|p_{0}|l u)\nonumber\\
&&-\frac{1}{2}\int_{0}^{1}du \frac{\sin (|p_{0}|l u)}{u} +\int_{0}^{1}du u
\sin (|p_{0}|lu)+\frac{1}{2}\int_{0}^{\infty}du \frac{\sin (u)}{u}\Big\},\nonumber
\end{eqnarray}
by adding and subtracting the term $\frac{1}{2}\int_{0}^{\infty}du \frac{\sin (u)}{u}$, and using $$\int_{0}^{\infty}du \frac{\sin (u)}{u}=\int_{0}^{\infty}du \frac{\sin (|p_{0}|l u)}{u}$$ for $|p_{0}|l\neq 0$. This expression shows that our $CPT$ violating term is characterized by the quantity  
\begin{eqnarray}\label{(21)}
\mu l^{2},
\end{eqnarray}
which has the dimension of mass.

For $|p_{0}|l\ll 1$ but $|p_{0}|\neq 0$ and using $\int_{0}^{\infty}du \frac{\sin (u)}{u}=\pi/2$, we have
\begin{eqnarray}\label{(22)}
f(p)
&\simeq&-\pi^{2}\mu l^{2}[\theta(p_{0})-\theta(-p_{0})]\theta(p^{2}),
\end{eqnarray}
which is Lorentz invariant. Note that the terms linear in $|p_{0}|l$ cancel out exactly although they do not appear to do so at first glance, and no infrared divergence in the $CPT$ violating term except for the subtlety in \eqref{(19)}. (Incidentally, if one chooses the non-local factor with an additional parameter $l^{\prime}$ in \eqref{(6)}, the factor $\pi^{2}\mu l^{2}$ is replaced by $\pi^{2}\mu (l^{2}-(l^{\prime})^{2})$ in \eqref{(22)}.)
 Thus the mass gap in \eqref{(17)} is given by 
\begin{eqnarray}\label{(23)}
\Delta m &\simeq&2\pi^{2}\mu l^{2},
\end{eqnarray}
provided that we adopt the prescription $ \sin(p_{0}L)=0 $
for $L\rightarrow\infty$ in \eqref{(19)}, which is natural in the sense of distribution $$\int dp_{0}\sin(p_{0}L) F(p_{0})=0,$$ for any test function $F(p_{0})$, by the Riemann-Lebesgue lemma.
 
Our definition of the non-local factor in \eqref{(6)} mostly eliminates the infrared singularity in the $CPT$ violating term except for the subtlety in \eqref{(19)}. Our $CPT$ violating term $f(p_{0})$ is odd in $p_{0}$ and 
$f(\pm 0)=\mp \Delta m/2$ but $f(0)=0$. The formula for the mass gap is defined for $|p_{0}|L\gg 1$ and thus precise $p_{0}=0$ is avoided.

As for the parity violating mass term in \eqref{(2.6)}, we have (in the frame with $\vec{p}=0$ for $p^{2}>0$)
\begin{eqnarray}\label{(24)}
g(p^{2})&=&\mu
\int d^{4}ze^{ipz}\left[\delta((z)^{2}-l^{2})-\delta((z)^{2})\right]\nonumber\\
&=&4\pi\mu\Big\{\int_{l}^{\infty}dy \sqrt{y^{2}-l^{2}}\cos [
p_{0}y]- \int_{0}^{\infty}dy y\cos [p_{0}y]\Big\},
\end{eqnarray}
which is evaluated as
\begin{eqnarray}\label{(25)}
g(p^{2})
&=&-4\pi\mu l^{2}
\Big\{\int_{1}^{\infty}du\frac{1}{\sqrt{u^{2}-1}+u}\cos [p_{0}lu] +\frac{\sin [p_{0}l]}{p_{0}l}+\frac{\cos [p_{0}l]-1}{(p_{0}l)^{2}}\Big\}.
\end{eqnarray}
This result also contains a subtlety as in \eqref{(19)} and mild logarithmic (or weaker) divergence at $p_{0}=0$ in the first term. This formula modulo a term as in \eqref{(19)} is again well-defined if precise $p_{0}=0$ is excluded.

\section{Neutrino-antineutrino mass splitting and baryogenesis}

The Lorentz invariant non-local factor 
\begin{eqnarray}\label{(26)}
\left[\theta(x^{0}-y^{0})-\theta(y^{0}-x^{0})\right]\delta\left((x-y)^{2}-l^{2}\right)
\end{eqnarray}
used in~\cite{CFT1, CFT2} induces infinite non-locality along the light-cone, and 
a dimensional counting shows that it diverges quadratically in the infrared
in momentum space. This infinite non-locality may lead to severe 
breaking of unitarity. In contrast, our modified  Lorentz invariant non-local factor   
\begin{eqnarray}\label{(27)}
\left[\theta(x^{0}-y^{0})-\theta(y^{0}-x^{0})\right]\left[\delta((x-y)^{2}-l^{2})-\delta((x-y)^{2}-(l^{\prime})^{2})\right],
\end{eqnarray}
(with $l^{\prime}=0$ in practical applications) mostly cancels out the infinite time-like volume effect and eliminates the quadratic
infrared divergence completely. In effect, our modified non-local factor induces non-locality which is limited within the fluctuation around the tip of the light-cone characterized by the length scale of the parameter $l$. This is very welcome from the point of view of unitarity, since
one can choose the length scale $l$ at the order of the Planck length and thus at least avoid the issue of unitarity violation in the framework of generic effective theory, which may be valid at a length scale much longer than the Planck scale.    
  
It is important to analyze if our mass splitting can induce a physically meaningful size of neutrino-antineutrino mass splitting by choosing $l$ of the order of Planck length and $\mu=M^{3}$ suitably, where $M$ is 
another mass scale of possible new physics. The natural neutrino mass splitting given by \eqref{(23)} is then
\begin{eqnarray}\label{(28)}
2\pi^{2} \mu l^{2}= 2\pi^{2}M(M/M_{P})^{2},
\end{eqnarray}
which is a kind of see-saw between $M$ and the Planck mass $M_{P}$. If one chooses $M\sim 10^{9}$ GeV, the mass splitting becomes of the order of the observed neutrino mass (difference) $\sim 0.1$ eV~\cite{particle-data}. This size of the mass scale $M$ is not uncommon in the leptogenesis based on the see-saw scenario~\cite{fukugita}, for example. 

If one assumes $M\sim 1000$ GeV, namely, the scale of the Standard Model, the mass splitting is $\sim 10^{-20}$ eV, which is too small
for  phenomenological interest. Although our model is for the neutrinos, the presently known experimental limit on the electron-positron mass splitting is~\cite{particle-data} 
\begin{eqnarray}\label{(29)}
\Delta m_{e}\leq 10^{-8}m_{e}\sim 10^{-2} eV.
\end{eqnarray}
To generate the corresponding value $\Delta m \sim 10^{-2}$ eV for the neutrinos in our scheme, we need a value slightly smaller than $M\sim 10^{9}$ GeV.

The neutrino mass splitting $\Delta m=10^{-1}\sim 10^{-2}$ eV, which is intended to be  of the order of $m_{D}/5$, is generated by $M\simeq 10^{8}\sim 10^{9}$ GeV and  appears to be allowed by presently available experimental data~\cite{pakvasa, adamson1}. 
Our Lorentz invariant model, which is written for the electron  but applicable to other leptons also, is concerned with neutrinos so far, but the higher order effect in renormalizable theory is generally expected to give rise to an electron-positron mass splitting  of the order $\Delta m_{e}\sim \alpha \Delta m$, with $\alpha$  the fine structure constant. This induced mass splitting, which is substantially smaller than $\Delta m=10^{-1} \sim 10^{-2}$ eV, is expected for all the massive charged leptons in the Standard Model. We, however, show in the next section that the induced effect is actually much smaller, $\sim 10^{-20}$ eV. Thus the value  $M\simeq 10^{8}\sim 10^{9}$ GeV is interesting for a phenomenological purpose, which does not apparently contradict any experimental data  and yet may be measurable in neutrino oscillation experiments in the near future. We tentatively adopt $\Delta m$ at  $10^{-1}\sim 10^{-2}$ eV, which is similar to the magnitude of $m_{D}$, in phenomenological discussions.

As for the baryogenesis, it is believed that leptons acquire masses from
electroweak symmetry breaking during the electroweak phase transition in the early universe. A neutrino-antineutrino mass difference
would then result in a leptonic  matter-antimatter asymmetry proportional to the mass difference. This asymmetry is transmitted to the baryon sector through the chiral anomaly~\cite{t-hooft} and  sphaleron processes~\cite{monton} which preserve $B - L$ but violate $B + L$. This "kinematical" picture implies the asymmetry
in the neutrino and antineutrino of the order~\cite{zeldovich} $(n_{\nu}-n_{\bar{\nu}})/ n_{\nu}\simeq m_{D}\Delta m/T^{2}$, which is, however,  too small at the electroweak energy scale to generate the baryon asymmetry via sphaleron processes in our case with  $\Delta m=10^{-1}\sim 10^{-2}$ eV. Besides, this initial asymmetry requires the lepton number non-conservation~\cite{zeldovich}, while the lepton number is conserved in our model without sphaleron effects.

Thus the lepton and quark sectors need to be treated simultaneously in the presence of sphalerons~\cite{monton} which break $B+L$ by preserving $B-L$. The authors of Ref.~\cite{barenboim} discuss a rather elaborate sphaleron dynamics by referring to ~\cite{rubakov,boguta,ringwald,dine}, and they conclude that the final baryon number at the energy scale of the weak mass $M_{W}$ is estimated at
\begin{eqnarray}\label{(30)}
\frac{n_{B}}{n_{\gamma}}\sim \frac{\Delta m}{M_{W}},
\end{eqnarray}
where $n_{\gamma}$ stands for the photon number density.
This estimate in the present case with $\Delta m=10^{-1}\sim 10^{-2}$ eV, namely $n_{B}/n_{\gamma}\sim 10^{-12} - 10^{-13}$, is smaller than the observed value $n_{B}/n_{\gamma} \simeq 10^{-10}$, but it still gives a promising number by considering the crude estimate in our model. The estimate of the generated baryon number is mainly constrained by experimental bounds on neutrino mass differences.

We emphasize that  {\em  this equilibrium electroweak baryogenesis does not need 
CP violation other than for the purpose of producing neutrino-antineutrino mass splitting.} The mechanism to generate \eqref{(30)} by a specific sphaleron dynamics~\cite{barenboim}, which is very interesting but requires further elaboration, differs from the more conventional baryogenesis~\cite{sakharov, yoshimura} and also from the leptogenesis~\cite{fukugita}. 

\section{Higher order induced effects}

We now analyze if the higher order effect due to the neutrino mass splitting is
well-controlled in the Standard Model. The propagator of the neutrino in the path integral formulation on the basis of Schwinger's action principle, which is based on the equations of motion, is given by~\cite{CFT2} (see also~\cite{fujikawa}), 
\begin{eqnarray}\label{(31)}
\langle T^{\star}\nu(x)\bar{\nu}(y)\rangle
=\int \frac{d^{4}p}{(2\pi)^{4}} e^{-ip(x-y)}\frac{i}{\pslash-m_{D} +i\epsilon+i\gamma_{5}g(p^{2})-i[f_{+}(p)-f_{-}(p)]},
\end{eqnarray}
where $f_{\pm}(p)$ and $g(p^{2})$ are defined in \eqref{(1.3)} and \eqref{(12)}, respectively, in 
connection with the free equation of motion \eqref{(2.6)}.  
  
We wish to examine the large momentum behavior of $f(p)=-i[f_{+}(p)-f_{-}(p)]$
defined in \eqref{(20)}
by assuming that $l$ is not very small and thus large $|p_{0}|l$ is physically relevant. 
By noting that $\frac{1}{2}\int_{0}^{1}du \frac{\sin (|p_{0}|l u)}{u}\rightarrow \frac{1}{2}\int_{0}^{\infty}du \frac{\sin (u)}{u}$
for $|p_{0}|l\rightarrow\infty$, the  expression \eqref{(20)} becomes
\begin{eqnarray}\label{(32)}
f(p)&=&-4\pi\mu l^{2}[\theta(p_{0})-\theta(-p_{0})]\theta(p^{2})\\
&&\times\Big\{\int_{1}^{\infty}du \frac{1}{2u(\sqrt{u^{2}-1}+u)^{2}}\sin (|p_{0}|l u)
+\frac{\sin (|p_{0}|l)}{(|p_{0}|l)^{2}}-\frac{\cos(|p_{0}|l)}{(|p_{0}|l)}\Big\}.\nonumber
\end{eqnarray}
The integral in this expression is shown to approach zero for  $|p_{0}|l\rightarrow\infty$ by the Riemann-Lebesgue lemma, and thus $f(p)$ approaches zero for  $|p_{0}|l\rightarrow\infty$.

The high energy behavior of the parity violating term $g(p^{2})$ in \eqref{(25)} is shown to be the same as the above $CPT$ violating term. 

The propagator \eqref{(31)} for Minkowski momentum is thus well behaved and the effects of non-locality are mild and limited
and, in this sense, $T^{\star}$ product may even be replaced by the canonical $T$ product in \eqref{(31)}~\cite{fujikawa}. (The logarithmic singularity in $g(p^{2})$ at $p_{\mu}=0$ may not be serious in the propagator \eqref{(31)} due to the presence of the phase space volume factor $d^{4}p$ in the numerator.)
In the analysis of the renormalization procedure, however, it is customary to consider the Euclidean amplitude obtained from the Minkowski amplitude by Wick rotation. This 
is because the power counting rule (superficial degree of divergence) of each Feynman amplitude is well defined with Euclidean momenta. Our propagator, which contains trigonometric functions, has undesirable behavior under the Wick rotation such as $\sin p_{0}z\rightarrow i\sinh p_{4}z$, and exponentially divergent behavior is generally induced and the effects of non-locality become significant. One might still argue that higher order effects in field theory defined on Minkowski space are in principle analyzed in Minkowski space and, if that is the case, our propagator maintains the ordinary renormalizable behavior.

One may assume that $l$ is of the order of Planck scale, as we did in our phenomenological analysis. Then  the higher order effect is expected to be small as long as $\mu l^{2}$ is small in either Minkowski or Euclidean formulation, since it is natural to assume that the loop momenta are cut-off below the Planck scale in generic effective theory and thus $|p_{0}|l$ may always be assumed to be small even inside the loop diagrams. In this case, our Lorentz invariant $CPT$ violating term is effectively replaced by
\begin{eqnarray}\label{(33)}
f(p)=-\pi^{2}\mu l^{2}[\theta(p_{0})-\theta(-p_{0})]\theta(p^{2}),
\end{eqnarray}
which is similar to a constant mass term except for the  $CPT$ violating 
factor $[\theta(p_{0})-\theta(-p_{0})] \theta(p^{2})$.
When this term is inserted into 
Feynman diagrams in the Standard Model, those Feynman diagrams are expected to show ordinary high energy behavior for a mass insertion, if the naive power counting works. 

\subsection*{Induced electron-positron mass splitting}
We now wish to show that the electron-positron mass splitting induced by the above factor $f(p)$ in \eqref{(33)}, when inserted into one-loop self-energy diagrams of the electron in the Standard Model, is well controlled. This correction amounts to the replacement of the ordinary momentum space neutrino propagator by 
\begin{eqnarray}\label{(34)}
\tilde{S}_{F}(p)_{\nu}=\frac{i}{\pslash-m_{D} +i\epsilon}f(p)\frac{i}{\pslash-m_{D}+i\epsilon}
\end{eqnarray}
in the one-loop self-energy diagrams of the electron in the Standard Model. We show that the electron self-energy correction induced by one-loop $W$-boson is well convergent, which implies that the momentum flowing through the neutrino propagator is limited by the $W$-boson mass scale $M_{W}$ and thus our starting assumption $|p|l\ll 1$ is justified.

\begin{figure*}[tbp]
\begin{center}
\includegraphics[scale=0.22]{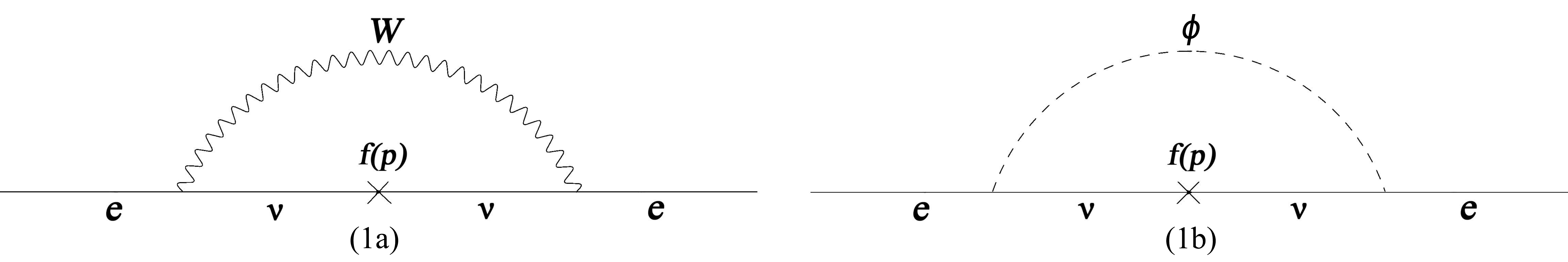}
\end{center}

\caption{Fig. (1a) represents the W-boson contribution and Fig. (1b) represents the charged unphysical Nambu--Goldstone $\phi$-boson contribution to the induced mass splitting between electron and positron.}
\label{fig1}
\end{figure*}

We work in the 't Hooft--Feynman gauge ($\xi=1$ in $R_{\xi}$-gauge~\cite{FLS} with Feynman rules given there). Then we have contributions from the two Feynman diagrams shown in Fig.1, with $\phi$ standing for the charged unphysical  Nambu--Goldstone boson.
The $W$-boson term in Fig.1a is then given by (ignoring precise numerical factors)   
\begin{eqnarray}\label{(35)}
&&g^{2}\int\frac{d^{4}p}{(2\pi)^{4}}\left[\gamma^{\alpha}\frac{(1-\gamma_{5})}{2}
\frac{\pslash+m_{D}}{p^{2}-m^{2}_{D} +i\epsilon}f(p)\frac{\pslash+m_{D}}{p^{2}-m^{2}_{D} +i\epsilon}\gamma_{\alpha}\frac{(1-\gamma_{5})}{2}\right]\nonumber\\
&&\times\frac{1}{(k-p)^{2}-M_{W}^{2} +i\epsilon},
\end{eqnarray}
where the numerator is proportional to $2m_{D}\pslash f(p)[(1-\gamma_{5})/2]$, and the integral is linearly convergent. This term is proportional to Dirac $\gamma^{\mu}$ and contributes to the electron kinetic term of the order $$\alpha [m_{D}\kslash/M_{W}^{2}](\mu l^{2})[\theta(k^{0})-\theta(-k^{0})] \theta(k^{2})[(1-\gamma_{5})/2],$$ where $\alpha$ is the fine structure constant. To infer this result, it is easier to analyze  \eqref{(35)} with $2m_{D}(kp) f(p)$ in the numerator by replacing $\gamma^{\mu}$ with $k^{\mu}$. It is then confirmed that the amplitude is invariant under the change of the
signature of $\vec{k}$ but changes the signature under the change of the signature of $k^{0}$ of the external electron at its rest frame $\vec{k}=0$. It  vanishes for space-like $k^{\mu}$, as is seen by choosing the frame with $k^{0}\rightarrow \pm0$, and it is Lorentz invariant. We thus infer $$\sim\alpha [m_{D}k^{2}/M_{W}^{2}](\mu l^{2})[\theta(k^{0})-\theta(-k^{0})] \theta(k^{2})$$ for \eqref{(35)} with
$2m_{D}(kp) f(p)$ in the numerator. One may finally replace $k^{2}$ by $\kslash$.

The charged $\phi$-boson term in Fig.1b is given by (ignoring precise numerical factors)
\begin{eqnarray}\label{(36)}
&&\frac{g^{2}}{M^{2}_{W}}\int\frac{d^{4}p}{(2\pi)^{4}}\left[m_{e}\frac{(1-\gamma_{5})}{2}-\frac{(1+\gamma_{5})}{2}m_{D}\right]
\frac{\pslash+m_{D}}{p^{2}-m^{2}_{D} +i\epsilon}f(p)\frac{\pslash+m_{D}}{p^{2}-m^{2}_{D} +i\epsilon}\nonumber\\
&&\times\left[m_{D}\frac{(1-\gamma_{5})}{2}-\frac{(1+\gamma_{5})}{2}m_{e}\right]
\frac{1}{(k-p)^{2}-M_{W}^{2} +i\epsilon},
\end{eqnarray}
where the deviation from the Standard Model result is given by the Lagrangian \eqref{standard}. The numerator contains two class of terms: the first class has the structure
$$[(m_{e}^{2}(1-\gamma_{5})/2+m_{D}^{2}(1+\gamma_{5})/2)/M_{W}^{2}]2m_{D}\pslash f(p),$$ which has a linearly convergent behavior as in the case of $W$-boson contribution and gives rise to a contribution of the order $$\alpha (1-\gamma_{5})/2[m_{e}^{2}/M_{W}^{2}][m_{D}\kslash/M_{W}^{2}](\mu l^{2})[\theta(k^{0})-\theta(-k^{0})] \theta(k^{2}).$$ The second class of terms contains in 
the numerator a factor $(m_{e}m_{D}/M_{W}^{2})(p^{2}+m^{2}_{D}) f(p)$ which potentially gives rise to a logarithmic divergence. If this divergence should persist, we need a new $CPT$ violating counter term for the electron, which would spoil the renormalizability of the Standard Model. But this term is in fact convergent due to 
the structure of the $CPT$ violating factor $[\theta(p_{0})-\theta(-p_{0})] \theta(p^{2})$, as 
\begin{eqnarray}\label{(37)}
&&\frac{g^{2}}{M^{2}_{W}}\int\frac{d^{3}p}{(2\pi)^{4}}\int_{0}^{\infty}dp_{0}\frac{(m_{e}m_{D})
(p^{2}+m_{D}^{2})}{(p^{2}-m^{2}_{D} +i\epsilon)^{2}}\theta(p^{2})(\mu l^{2})\nonumber\\
&&\times\left[\frac{1}{(k-p)^{2}-M_{W}^{2} +i\epsilon}-\frac{1}{(k+p)^{2}-M_{W}^{2} +i\epsilon}\right].
\end{eqnarray}  
This is linearly convergent, and it is independent of the Dirac $\gamma^{\mu}$, thus contributing to the mass correction. Remark that this term
changes its signature under the change of the signature of $k^{0}$ of the external electron at its rest frame $\vec{k}=0$ and vanishes for the space-like $k^{\mu}$ as is seen by choosing the frame with $k^{0}\rightarrow \pm0$. This term also vanishes at $k^{\mu}=0$, and it is estimated at the order of
$$\alpha (m_{e}m_{D}/M_{W}^{2})(k^{2}/M_{W}^{2})(\mu l^{2})[\theta(k^{0})-\theta(-k^{0})] \theta(k^{2}),$$
which is $$\alpha (m_{e}m_{D}/M_{W}^{2})(m_{e}^{2}/M_{W}^{2})(\mu l^{2})[\theta(k^{0})-\theta(-k^{0})] \theta(k^{2})$$ near on-shell.  

The induced effect on the electron is determined by the eigenvalue equation (by ignoring ordinary corrections in the Standard Model, for simplicity)
\begin{eqnarray}\label{(38)}
\kslash (1+\alpha A(k))-(m_{e}+\alpha B(k))=0
\end{eqnarray}
or equivalently $\kslash= m_{e}+\alpha B(k) - m_{e}\alpha A(k)$, where $A(k)$ and $B(k)$
stand for the induced effects evaluated above. The induced $CPT$ violating effect on the electron- positron splitting is thus finite and the leading contribution is given by the $W$-boson at the order,
\begin{eqnarray}\label{(39)}
\alpha [m_{D}m_{e}/M_{W}^{2}](\mu l^{2})[(1-\gamma_{5})/2][\theta(k^{0})-\theta(-k^{0})] \theta(k^{2}),
\end{eqnarray}
which, if we choose our parameter at $\pi^{2}\mu l^{2}=10^{-1}\sim 10^{-2}$ eV as in the previous section, is $\sim 10^{-20}$ eV and thus well below the present experimental bound. The induced $CPT$ violation is expected to be smaller in the quark sector (as a two-loop effect) than in the charged leptons in the $SU(2)\times U(1)$ invariant theory, and thus much smaller than the well-known limit on the $K$-meson, $|m_{K}-m_{\bar{K}}|<0.44\times 10^{-18}$ GeV~\cite{particle-data}.

\section{Conclusion}
We have demonstrated that a well-defined neutrino-antineutrino mass splitting can be realized in the Standard Model. It has potentially interesting implications on equilibrium electroweak baryogenesis, depending on the subtle details of sphaleron dynamics. The Lorentz invariant $CPT$ breaking mass term in momentum space is effectively represented by 
$f(p)=-(\Delta m/2)[\theta(p_{0})-\theta(-p_{0})]\theta(p^{2})$.
The induced  $CPT$ violating effect on the electron-positron mass splitting in the Standard Model is shown to be finite and kept well below the experimental bound.
Our analysis shows that non-local Lorentz invariant $CPT$ violation is very natural in the Standard Model and it would be interesting to search for possible neutrino-antineutrino mass splitting in oscillation experiments. A physical origin  of our Planck scale non-local $CPT$ violation, which is most likely related to quantum gravity, remains to be clarified.

\section*{Acknowledgments}
We thank 
M. Chaichian for stimulating discussions. We also thank 
R. Bufalo for drawing the figures. The support of the Academy of Finland under the Projects No. 136539 and 272919, of the Vilho, Yrj\"o and Kalle V\"ais\"al\"a Foundation, and of JSPS KAKENHI (Grant No. 25400415) is gratefully acknowledged.

\end{document}